\documentclass[12pt]{article}
\usepackage{bbm}
\usepackage{amssymb}
\usepackage{pifont}
\usepackage{mathrsfs}
\usepackage{amsfonts}
\usepackage{color}
\usepackage{footmisc}
\usepackage{cite}

\newcommand{\omits}[1]{}

\begin{document}

\begin{center}
{\bf \Large Cosmology of a polynomial model for de Sitter\bigskip

gauge theory sourced by a fluid}
\bigskip\bigskip

{\large Jia-An Lu\footnote{Email: ljagdgz@163.com}}
\bigskip

School of Physics, Sun Yat-sen University,\\ Guangzhou 510275, China
\bigskip

\begin{abstract}
In the de Sitter gauge theory (DGT), the fundamental variables are the de Sitter (dS) connection and
the gravitational Higgs/Goldstone field $\xi^A$.
Previously, a model for DGT was analyzed, which generalizes the MacDowell--Mansouri
gravity to have a variable cosmological constant $\Lambda=3/l^2$, where $l$ is related to $\xi^A$
by $\xi^A\xi_A=l^2$.
It was shown that the model sourced by a perfect fluid does not support a radiation epoch and the
accelerated expansion of the parity invariant universe.
In this work, I consider a similar model, namely, the Stelle--West gravity, and couple it to a
modified perfect fluid, such that the total Lagrangian 4-form is polynomial in the gravitational variables.
The Lagrangian of the modified fluid has a nontrivial variational derivative with respect to $l$,
and as a result, the problems encountered in the previous work no longer appear.
Moreover, to explore the elegance of the general theory, as well as to write down the basic framework,
I perform the Lagrange--Noether analysis for DGT sourced by a matter field, yielding the
field equations and the identities with respect to the symmetries of the system. The resulted formula
are dS covariant and do not rely on the existence of the metric field.
\end{abstract}
\end{center}



\section{Introduction}
The gauge theories of gravity (GTG) aim at treating gravity as a gauge field, in particular,
constructing a Yang--Mills-type Lagrangian which reduces to GR in some limiting case, while
providing some novel falsifiable predictions. A well-founded subclass of GTG is the Poincar\'e
gauge theory (PGT) \cite{Kibble,Sciama,Hehl13,Obukhov17,Mielke}, in which the gravitational field
consists of the Lorentz connection and the co-tetrad field. Moreover, the PGT can be reformulated
as de Sitter gauge theory (DGT), in which the Lorentz connection and the co-tetrad field are united
into a de Sitter (dS) connection \cite{West79,West80}. In fact, before the idea of DGT is realized,
a related Yang--Mills-type Lagrangian for gravity was proposed by MacDowell and Mansouri \cite{MM},
and reformulated into a dS-invariant form by West \cite{West78}, which reads
\begin{eqnarray}\label{MM}
{\cal L}^{\rm MM}&=&\epsilon_{ABCDE}\,\xi^E{\cal F}^{AB}\wedge{\cal F}^{CD}\nonumber\\
&=&\epsilon_{\alpha\beta\gamma\delta}\,(lR^{\alpha\beta}\wedge R^{\gamma\delta}
-2l^{-1}R^{\alpha\beta}\wedge e^\gamma\wedge e^\delta
+l^{-3}e^\alpha\wedge e^\beta\wedge e^\gamma\wedge e^\delta),
\end{eqnarray}
where $\epsilon_{ABCDE}$ and $\epsilon_{\alpha\beta\gamma\delta}$ are the 5d and 4d Levi-Civita symbols,
$\xi^A$ is a dS vector constrained by $\xi^A\xi_A=l^2$, $l$ is a positive constant, ${\cal F}^{AB}$
is the dS curvature, $R^{\alpha\beta}$ is the Lorentz curvature, and $e^\alpha$
is the orthonormal co-tetrad field. This theory is equivalent to the Einstein--Cartan (EC) theory
with a cosmological constant $\Lambda=3/l^2$ and a Gauss--Bonnet (GB) topological term, as seen
in Eq. (\ref{MM}).

Note that some special gauges with the residual Lorentz symmetry can be defined
by $\xi^A=\delta^A{}_4l$. Henceforth, $\xi^A$ is akin to an unphysical Goldstone field. To make
$\xi^A$ physical, and so become the gravitational Higgs field, one may replace the constant $l$
by a dynamical $l$, resulting in the Stelle--West (SW) theory \cite{West80}. The theory is further
explored by Refs. \cite{Westman14,Magueijo} (see also the review \cite{Westman15}), in which
the constraint $\xi^A\xi_A=l^2$ is completely removed, in other words,
$\xi^A\xi_A$ needs not to be positive. Suppose that $\xi^A\xi_A=\sigma l^2$, where $\sigma=\pm1$.
When $l\neq0$, the metric field can be defined by $g_{\mu\nu}=(\widetilde{D}_\mu\xi^A)(\widetilde{D}_\nu
\xi_A)$, where $\widetilde{D}_\mu\xi^A=\widetilde{\delta}^A{}_BD_\mu\xi^B$, $\widetilde{\delta}^A{}_B=
\delta^A{}_B-\xi^A\xi_B/\sigma l^2$, $D_\mu\xi^A=d_\mu\xi^A+\Omega^A{}_{B\mu}\xi^B$, and $\Omega^A{}
_{B\mu}$ is the dS connection. It was shown that $\sigma=\pm1$ corresponds to the Lorentz/Euclidean
signature of the metric field, and the signature changes when $\xi^A\xi_A$ changes its sign \cite{Magueijo}.

On the other hand, it remains to check whether the SW gravity is viable. Although the SW lagrangian
reduces to the MM Lagrangian when $l$ is a constant, the field equations do not. In the SW theory,
there is an additional field equation coming from the variation with respect to $l$, which is nontrivial
even when $l$ is a constant. Actually, a recent work \cite{Alexander} presents some negative results for
a related model, whose Lagrangian is equal to the SW one times $(-l/2)$. For a homogeneous and isotropic
universe with parity-invariant torsion, it is found that $l$ being a constant implies the energy density
of the material fluid being a constant, and so $l$ should not be a constant in the
general case. Moreover, in the radiation epoch, the $l$ equation forces the energy density equal to zero;
while
in the matter epoch, a dynamical $l$ only works to renormalize the gravitational constant by some
constant factor, and hence the cosmic expansion decelerates as in GR.

In this work, it is shown that the SW gravity suffers from similar problems encountered in the model
considered by Ref. \cite{Alexander}. Also, I try to solve these problems by using a new fluid with the
Lagrangian being polynomial in the gravitational variables. The merits of a Lagrangian polynomial
in some variables are that it is simple and nonsingular with respect to those variables.
In Refs. \cite{Pagels,Westman13}, the polynomial Lagrangian for gravitation and other fundamental fields
were proposed, while in this paper, the polynomial Lagrangian for an perfect fluid is proposed, which
reduces to the Lagrangian of a usual perfect fluid when $l$ is a constant. It turns out that, in contrast to the
case with an ordinary fluid, the SW gravity coupled with the new fluid supports the radiation epoch and
naturally drives the cosmic
acceleration. In addition, when writing down the basic framework of DGT, a Lagrangian--Noether analysis
is performed, which generalizes the results of Ref. \cite{Lu16} to the cases with arbitrary matter field
and arbitrary $\xi^A$.

The article is organized as follows. In Sec. \ref{2.1}, a Lagrangian--Noether analysis
is done for the general DGT sourced by a matter field. In Sec. \ref{2.2}, I reduce the analysis in
Sec. \ref{2.1} in the Lorentz gauges, and show how the two Noether identities in PGT can be elegantly
unified into one identity in DGT. In Sec. \ref{3.1}, the SW model of DGT is introduced, with the field
equations derived both in the general gauge and the Lorentz gauges. Further, the matter source is discussed
in Sec. \ref{3.2}, where a modified perfect fluid with the Lagrangian polynomial in the gravitational
variables is constructed, and a general class of perfect fluids is defined, which contains both the
usual and the modified perfect fluids. Then I couple the SW gravity with the class of fluids and study
the coupling system in the homogeneous, isotropic and parity-invariant universe. The field equations
are deduced in Sec. \ref{4.1} and solved in Sec. \ref{4.2}, and the results are compared with observations
in Sec. \ref{4.3}. In Sec. \ref{5}, I give some conclusions, and discuss the remaining problems and possible
solutions.

\section{de Sitter gauge theory}
\subsection{Lagrangian--Noether machinery}\label{2.1}
The DGT sourced by a matter field is described by the Lagrangian 4-form
\begin{equation}\label{L}
{\cal L}={\cal L}(\psi,D\psi,\xi^A,D\xi^A,{\cal F}^{AB}),
\end{equation}
where $\psi$ is a $p$-form valued at some representation space of the dS group $SO(1,4)$, $D\psi=d\psi+
\Omega^{AB}T_{AB}\wedge\psi$ is the covariant exterior derivative, $T_A{}^B$ are representations of the
dS generators, $\xi^A$ is a dS vector, $D\xi^A=d\xi^A+\Omega^A{}_B\xi^B$,
$\Omega^A{}_B$ is the dS connection 1-form, and ${\cal F}^A{}_B=d\Omega^A{}_B+\Omega^A{}_C
\wedge\Omega^C{}_B$ is the dS curvature 2-form. The variation of ${\cal L}$ resulted from the variations
of the explicit variables reads
\begin{eqnarray}\label{dL}
\delta{\cal L}&=&\delta\psi\wedge\partial{\cal L}/\partial\psi+\delta D\psi\wedge\partial{\cal L}/\partial D\psi
+\delta\xi^A\cdot\partial{\cal L}/\partial\xi^A+\delta D\xi^A\wedge\partial{\cal L}/\partial D\xi^A\nonumber\\
&{}&+\delta {\cal F}^{AB}\wedge\partial{\cal L}/\partial {\cal F}^{AB},
\end{eqnarray}
where $(\partial{\cal L}/\partial\psi)_{\mu_{p+1}\cdots\mu_4}\equiv\partial{\cal L}_{\mu_1\cdots\mu_p
\mu_{p+1}\cdots\mu_4}/\partial\psi_{\mu_1\cdots\mu_p}$, and the other partial derivatives
are similarly defined. The variations of $D\psi$, $D\xi^A$ and ${\cal F}^{AB}$ can be transformed into
the variations of the fundamental variables $\psi$, $\xi^A$, and $\Omega^{AB}$, leading to
\begin{eqnarray}\label{dL2}
\delta{\cal L}&=&\delta\psi\wedge V_\psi+\delta\xi^A\cdot V_A+\delta\Omega^{AB}\wedge V_{AB}\nonumber\\
&{}&+d(\delta\psi\wedge\partial{\cal L}/\partial D\psi+\delta\xi^A\cdot\partial{\cal L}/\partial D\xi^A
+\delta\Omega^{AB}\wedge\partial{\cal L}/\partial{\cal F}^{AB}),
\end{eqnarray}
where
\begin{equation}
V_\psi\equiv\delta{\cal L}/\delta\psi=\partial{\cal L}/\partial\psi-(-1)^pD\partial{\cal L}/\partial D\psi,
\end{equation}
\begin{equation}\label{VA}
V_A\equiv\delta{\cal L}/\delta\xi^A=\partial{\cal L}/\partial\xi^A-D\partial{\cal L}/\partial D\xi^A,
\end{equation}
\begin{equation}\label{VAB}
V_{AB}\equiv\delta{\cal L}/\delta\Omega^{AB}=T_{AB}\psi\wedge\partial{\cal L}/\partial D\psi
+\partial{\cal L}/\partial D\xi^{[A}\cdot\xi_{B]}+D\partial{\cal L}/\partial {\cal F}^{AB}.
\end{equation}

The symmetry transformations in DGT consist of the diffeomorphism transformations and the dS
transformations. For the diffeomorphism transformations, they can be promoted to a gauge-invariant
version \cite{Hehl76,Lu16}, namely, the parallel transports in the fiber bundle with the gauge group as
the structure group. The action of an infinitesimal parallel transport on a variable is a
gauge-covariant Lie derivative\,\footnote{The gauge-covariant Lie derivative has been used in the
metric-affine gauge theory of gravity \cite{Hehl95}.} $L_v\equiv v\rfloor D+Dv\rfloor$, where $v$ is the
vector field which generates the infinitesimal parallel transport,
and $\rfloor$ denotes a contraction, for example, $(v\rfloor \psi)_{\mu_2\cdots\mu_p}=v^{\mu_1}
\psi_{\mu_1\mu_2\cdots\mu_p}$. Put $\delta=L_v$ in Eq. (\ref{dL}), utilize the arbitrariness of $v$, then
one obtains the chain rule
\begin{eqnarray}
v\rfloor{\cal L}&=&(v\rfloor\psi)\wedge\partial{\cal L}/\partial\psi+(v\rfloor D\psi)\wedge\partial{\cal L}
/\partial D\psi+(v\rfloor D\xi^A)\cdot\partial{\cal L}/\partial D\xi^A\nonumber\\
&{}&+(v\rfloor {\cal F}^{AB})\wedge\partial{\cal L}/\partial {\cal F}^{AB},
\end{eqnarray}
and the first Noether identity
\begin{equation}\label{N1}
(v\rfloor D\psi)\wedge V_\psi+(-1)^p(v\rfloor\psi)\wedge DV_\psi+(v\rfloor D\xi^A)\cdot V_A
+(v\rfloor{\cal F}^{AB})\wedge V_{AB}=0.
\end{equation}
On the other hand, the dS transformations are defined as vertical isomorphisms on the fiber bundle.
The actions of an infinitesimal dS transformation on the fundamental variables are as follows:
\begin{equation}\label{dSt}
\delta\psi=B^{AB}T_{AB}\psi,\quad \delta\xi^A=B^{AB}\xi_B,\quad \delta\Omega^{AB}=-DB^{AB},
\end{equation}
where $B^A{}_B$ is a dS algebra-valued function which generates the infinitesimal dS transformation.
Substitute Eq. (\ref{dSt}) and $\delta{\cal L}
=0$ into Eq. (\ref{dL2}), and make use of Eq. (\ref{VAB}) and the arbitrariness of $B^{AB}$, one arrives
at the second Noether identity
\begin{equation}\label{N2}
DV_{AB}=-T_{AB}\psi\wedge V_\psi-V_{[A}\cdot\xi_{B]}.
\end{equation}

The above analyses are so general that they do not require the existence of a metric field. In the special
case with a metric field being defined, $\xi^A\xi_A$ equating to a positive constant, and $p=0$, the above
analyses coincide with those in Ref. \cite{Lu16}.

\subsection{Reduction in the Lorentz gauges}\label{2.2}
Consider the case with $\xi^A\xi_A=l^2$, where $l$ is a positive function. Then we may define
the projector $\widetilde{\delta}^A{}_B=\delta^A{}_B-\xi^A\xi_B/l^2$, the generalized tetrad
$\widetilde{D}\xi^A=\widetilde{\delta}^A{}_B D\xi^B$, and a symmetric rank-2 tensor\,\footnote{This
formula has been given by Refs. \cite{Lu13,Magueijo}, and is different from that originally
proposed by Stelle and West \cite{West80} by a factor $(l_0/l)^2$, where $l_0$ is the vacuum expectation
value of $l$.}
\begin{equation}\label{gmn}
g_{\mu\nu}=\eta_{AB}(\widetilde{D}_\mu\xi^A)(\widetilde{D}_\nu\xi^B),
\end{equation}
which is a localization of the dS metric $\mathring{g}_{\mu\nu}=\eta_{AB}(d_\mu\mathring{\xi}^A)
(d_\nu\mathring{\xi}^B)$, where $\mathring{\xi}^A$ are the 5d Minkowski coordinates on the 4d dS space.
Though Eq. (\ref{gmn}) seems less natural than the choice $g^*_{\mu\nu}=\eta_{AB}(D_\nu\xi^A)
(D_\nu\xi^B)$, it coincides with another natural identification (\ref{Omega}) (the relation between Eqs.
(\ref{gmn}) and (\ref{Omega}) will be discussed later).
If $g_{\mu\nu}$ is non-degenerate, it is a metric field with Lorentz signature, and one may define
$\widetilde{D}^\mu\xi_A\equiv g^{\mu\nu}\widetilde{D}_\nu\xi_A$. Put $v^\mu=\widetilde{D}
^\mu\xi_A$ in Eq. (\ref{N1}) and utilize
$(\widetilde{D}_\mu\xi^A)(\widetilde{D}^\mu\xi_B)=\widetilde{\delta}^A{}_B$, we get
\begin{eqnarray}\label{VtA}
\widetilde{V}_A&=&-(\widetilde{D}\xi_A\rfloor D\psi)\wedge V_\psi
-(-1)^p(\widetilde{D}\xi_A\rfloor\psi)\wedge DV_\psi
-(\widetilde{D}\xi_A\rfloor d\ln l)\cdot V_C\xi^C\nonumber\\
&{}&-(\widetilde{D}\xi_A\rfloor{\cal F}^{CD})\wedge V_{CD},
\end{eqnarray}
where $\widetilde{V}_A=\widetilde{\delta}^B{}_AV_B$. When $l$ is a constant, Eq. (\ref{VtA}) implies
that the $\xi^A$ equation ($\widetilde{V}_A=0$ for this case) can be deduced from the other field
equations ($V_\psi=0$ and $V_{CD}=0$), as pointed out by Ref. \cite{Lu13}. Substitute Eq. (\ref{VtA})
into Eq. (\ref{N2}), and make use of $\widetilde{V}_{[A}\cdot\xi_{B]}=V_{[A}\cdot\xi_{B]}$ and
$\widetilde{D}\xi_{[A}\cdot\xi_{B]}=D\xi_{[A}\cdot\xi_{B]}$, one attains
\begin{eqnarray}\label{DV2}
DV_{AB}&=&-T_{AB}\psi\wedge V_\psi+(D\xi_{[A}\cdot\xi_{B]}\rfloor D\psi)\wedge V_\psi
+(-1)^p(D\xi_{[A}\cdot\xi_{B]}\rfloor \psi)\wedge DV_\psi\nonumber\\
&{}&+(D\xi_{[A}\cdot\xi_{B]}\rfloor d\ln l)\cdot V_C\xi^C
+(D\xi_{[A}\cdot\xi_{B]}\rfloor {\cal F}^{CD})\wedge V_{CD}.
\end{eqnarray}
When $l$ is a constant, Eq. (\ref{DV2}) coincides with the corresponding result in Ref. \cite{Lu16}.
As will be shown later, Eq. (\ref{DV2}) unifies the two Noether identities in PGT.

To see this, let us define the Lorentz gauges by the condition $\xi^A=\delta^A{}_4l$ \cite{West80}. If
$h^A{}_B$ $\in SO(1,4)$ preserves these gauges, then $h^A{}_B={\rm diag}(h^\alpha{}_\beta,1)$, where $h^\alpha
{}_\beta$ belongs to the Lorentz group $SO(1,3)$. In the Lorentz gauges, $\Omega^\alpha{}_\beta$ transforms
as a Lorentz connection, and $\Omega^\alpha{}_4$ transforms as a co-tetrad field. Therefore, one may identify
$\Omega^\alpha{}_\beta$ as the spacetime connection $\Gamma^\alpha{}_\beta$, and $\Omega^\alpha{}_4$ as the
co-tetrad field $e^\alpha$ divided by some quantity with the dimension of length, a natural choice for which
is $l$. Resultantly, $\Omega^{AB}$ is identified with a combination of geometric quantities as follows:
\begin{equation}\label{Omega}
\Omega^{AB}=\left(
\begin{array}{cc}
\Gamma^{\alpha\beta}&l^{-1}e^{\alpha}\\
-l^{-1}e^\beta&0
\end{array}
\right).
\end{equation}
In the case with constant $l$, this formula has been given by Refs. \cite{Guo76,West80}, and, in the
case with varying $l$, it has been given by Refs. \cite{Lu13,Westman14}. In the Lorentz
gauges, $\widetilde{D}\xi^4=0$, $\widetilde{D}\xi^\alpha=\Omega^\alpha{}_4l=e^\alpha$
(where Eq. (\ref{Omega}) is used), and so $g_{\mu\nu}$
defined by Eq. (\ref{gmn}) satisfies $g_{\mu\nu}=\eta_{\alpha\beta}e^\alpha{}_\mu e^\beta{}_\nu$,
implying that Eq. (\ref{gmn}) coincides with Eq. (\ref{Omega}). Moreover, according to Eq. (\ref{Omega}),
one finds the expression for ${\cal F}^{AB}$ in the Lorentz gauges as follows \cite{Lu13}:
\begin{equation}\label{F}
\mathcal {F}^{AB}=\left(
\begin{array}{cc}
R^{\alpha\beta}-l^{-2}e^{\alpha}\wedge e^\beta
&l^{-1}[S^{\alpha}-d\ln l\wedge e^\alpha]\\
-l^{-1}[S^\beta-d\ln l\wedge e^\beta]&0
\end{array}
\right),
\end{equation}
where $R^\alpha{}_\beta=d\Gamma^\alpha{}_\beta+\Gamma^\alpha{}_\gamma\wedge\Gamma^\gamma{}_\beta$ is the
spacetime curvature, and $S^\alpha=de^\alpha+\Gamma^\alpha{}_\beta\wedge e^\beta$ is the spacetime torsion.

Now it is ready to interpret the results in Sec. \ref{2.1} in the Lorentz gauges. In those gauges,
$D\psi=D^\Gamma\psi+2l^{-1}e^\alpha T_{\alpha4}\wedge\psi$, $D\xi^\alpha=e^\alpha$, $D\xi^4=dl$,
and so Eq. (\ref{L}) becomes
\begin{equation}
{\cal L}={\cal L}^L(\psi,D^\Gamma\psi,l,dl,e^\alpha,R^{\alpha\beta},S^\alpha),
\end{equation}
where $D^\Gamma\psi=d\psi+\Gamma^{\alpha\beta} T_{\alpha\beta}\wedge\psi$. It is the same as a
Lagrangian 4-form in PGT \cite{Obukhov}, with the fundamental variables being $\psi$, $l$,
$\Gamma^{\alpha\beta}$
and $e^\alpha$. The relations between the variational derivatives with respect to the PGT variables
and those with respect to the DGT variables can be deduced from the following equality:
\begin{equation}
\delta\xi^A\cdot V_A+2\delta\Omega^{\alpha4}\wedge V_{\alpha4}
=\delta l\cdot \Sigma_l+\delta e^\alpha\wedge \Sigma_\alpha,
\end{equation}
where $\Sigma_l\equiv\delta{\cal L}^L/\delta l$ and $\Sigma_\alpha\equiv\delta{\cal L}^L/\delta e^\alpha$.
Explicitly, the relations are:
\begin{equation}\label{Spsi}
\Sigma_\psi\equiv\delta{\cal L}^L/\delta\psi=V_\psi,
\end{equation}
\begin{equation}
\Sigma_l=V_4-2l^{-2}e^\alpha\wedge V_{\alpha4},
\end{equation}
\begin{equation}
\Sigma_{\alpha\beta}\equiv\delta{\cal L}^L/\delta\Gamma^{\alpha\beta}=V_{\alpha\beta},
\end{equation}
\begin{equation}\label{Salpha}
\Sigma_\alpha=2l^{-1}V_{\alpha4}.
\end{equation}
It is remarkable that the DGT variational derivative $V_{AB}$ unifies the two PGT variational derivatives
$\Sigma_{\alpha\beta}$ and $\Sigma_\alpha$. With the help of Eqs. (\ref{Spsi})--(\ref{Salpha}), the
$\alpha\beta$ components and $\alpha4$ components of Eq. (\ref{DV2}) are found to be
\begin{equation}
D^\Gamma\Sigma_{\alpha\beta}=-T_{\alpha\beta}\psi\wedge\Sigma_\psi+e_{[\alpha}\wedge\Sigma_{\beta]},
\end{equation}
\begin{eqnarray}
D^\Gamma\Sigma_\alpha&=&D^\Gamma_\alpha\psi\wedge\Sigma_\psi+(-1)^p(e_\alpha\rfloor\psi)\wedge D^\Gamma\Sigma_\psi
+\partial_\alpha l\cdot\Sigma_l\nonumber\\
&{}&+(e_\alpha\rfloor R^{\beta\gamma})\wedge\Sigma_{\beta\gamma}+(e_\alpha\rfloor S^\beta)\wedge\Sigma_\beta,
\end{eqnarray}
which are just the two Noether identities in PGT \cite{Obukhov}, with both $\psi$ and $l$ as the matter fields.
This completes
our proof for the earlier statement that the DGT identity (\ref{DV2}) unifies the two Noether identities in PGT.

\section{Polynomial models for DGT}
\subsection{Stelle--West gravity}\label{3.1}
It is natural to require that the Lagrangian for DGT is regular with respect to the fundamental variables.
The simplest regular Lagrangian are polynomial in the variables, and, in order to recover the EC theory,
the polynomial Lagrangian should be at least linear in the gauge curvature. Moreover, to ensure
${\cal F}^{AB}=0$ is naturally a vacuum solution, the polynomial Lagrangian should be at least quadratic in
${\cal F}^{AB}$\,\footnote{When the Lagrangian is linear in ${\cal F}^{AB}$, we may add some `constant term'
(independent of ${\cal F}^{AB}$) to ensure ${\cal F}^{AB}=0$ is a vacuum solution, but this way is not so
natural.}. The general Lagrangian quadratic in ${\cal F}^{AB}$ reads:
\begin{eqnarray}\label{LG}
{\cal L}^{\rm G}&=&(\kappa_1\,\epsilon_{ABCDE}\,\xi^E+\kappa_2\,\eta_{AC}\xi_B\xi_D+\kappa_3\,\eta_{AC}
\eta_{BD}){\cal F}^{AB}\wedge{\cal F}^{CD}\nonumber\\
&=&\kappa_1{\cal L}^{\rm SW}+\kappa_2(S^\alpha\wedge S_\alpha
-2S^\alpha\wedge d\ln l\wedge e_\alpha)\nonumber\\
&{}&+\kappa_3[R^{\alpha\beta}\wedge R_{\alpha\beta}+d(2l^{-2}S^\alpha\wedge e_\alpha)],
\end{eqnarray}
where the $\kappa_1$ term is the SW Lagrangian, the $\kappa_2$ and $\kappa_3$ terms are parity odd, and the
$\kappa_3$ term is a sum of the Pontryagin and modified Nieh--Yan topological terms. This quadratic Lagrangian
is a special case of the at most quadratic Lagrangian proposed in Refs. \cite{Westman12,Westman14}, and one should note
that the quadratic Lagrangian satisfies the requirement mentioned above about the vacuum solution, while the
at most quadratic Lagrangian does not always satisfy that requirement.

Among the three terms in Eq. (\ref{LG}), the SW term is the only one that can be reduced to the EC Lagrangian
in the case with positive and constant $\xi^A\xi_A$. Thus the SW Lagrangian is the simplest choice for the
gravitational Lagrangian which (i) is regular with respect to the fundamental variables; (ii) can be reduced
to the EC Lagrangian; (iii) ensures ${\cal F}^{AB}=0$ is naturally a vacuum solution.

The SW Lagrangian 4-form ${\cal L}^{\rm SW}$ takes the same form as ${\cal L}^{\rm MM}$ in the first
line of Eq. (\ref{MM}),
while $\xi^A$ is not constrained by any condition. Substitute Eq. (\ref{MM})
into Eqs. (\ref{VA})--(\ref{VAB}), make use of $\partial{\cal L}^{\rm SW}/\partial{\cal F}^{AB}=\epsilon_{ABCDE}
\,\xi^E{\cal F}^{CD}$ and the Bianchi identity $D{\cal F}^{AB}=0$, one immediately gets the gravitational
field equations
\begin{equation}\label{Eq1}
-\kappa\,\epsilon_{ABCDE}\,{\cal F}^{AB}\wedge{\cal F}^{CD}=\delta{\cal L}^{\rm m}/\delta\xi^E,
\end{equation}
\begin{equation}\label{Eq2}
-\kappa\,\epsilon_{ABCDE}\,D\xi^E\wedge{\cal F}^{CD}=\delta{\cal L}^{\rm m}/\delta\Omega^{AB},
\end{equation}
where ${\cal L}^{\rm m}$ is the Lagrangian of the matter field coupled to the SW gravity, with
$\kappa$ as the coupling constant. In the vacuum case, Eq. (\ref{Eq2}) has been given by Ref. \cite{Westman12}
by direct computation, while here, Eq. (\ref{Eq2}) is obtained from the general formula (\ref{VAB}).

In the Lorentz gauges, ${\cal L}^{\rm SW}$ takes the same form as ${\cal L}^{\rm MM}$ in the second
line of Eq. (\ref{MM}), while $l$ becomes a dynamical field. The gravitational field equations read
\begin{equation}\label{eq-l}
-(\kappa/4)\epsilon_{\alpha\beta\gamma\delta}\,\epsilon^{\mu\nu\sigma\rho}e^{-1}R^{\alpha\beta}{}_{\mu\nu}
R^{\gamma\delta}{}_{\sigma\rho}-4\kappa\,l^{-2}R+72\kappa\,l^{-4}=\delta S_{\rm m}/\delta l,
\end{equation}
\begin{equation}\label{eq-Gamma}
-\kappa\,\epsilon_{\alpha\beta\gamma\delta}\,\epsilon^{\mu\nu\sigma\rho}e^{-1}\partial_\nu l\cdot
R^{\gamma\delta}{}_{\sigma\rho}+8\kappa\,e_{[\alpha}{}^\mu e_{\beta]}{}^\nu\partial_\nu l^{-1}
+4\kappa\,l^{-1}T^\mu{}_{\alpha\beta}=\delta S_{\rm m}/\delta\Gamma^{\alpha\beta}{}_\mu,
\end{equation}
\begin{equation}\label{eq-e}
-8\kappa\,l^{-1}(G^\mu{}_\alpha+\Lambda e_\alpha{}^\mu)=\delta S_{\rm m}/\delta e^\alpha{}_\mu,
\end{equation}
where $e=\det(e^\alpha{}_\mu)$, $R$ is the scalar curvature, $G^\mu{}_\alpha$ is the Einstein tensor,
$T^\mu{}_{\alpha\beta}=S^\mu{}_{\alpha\beta}+2e_{[\alpha}{}^\mu S^\nu{}_{\beta]\nu}$, and $S_{\rm m}$
is the action of the matter field.

\subsection{Polynomial dS fluid}\label{3.2}
For the same reason of choosing a polynomial Lagrangian for DGT, we intend to use those matter sources
with polynomial Lagrangian. It has been shown that the Lagrangian of fundamental fields can be reformulated
into polynomial forms \cite{Pagels,Westman13}. However, when describing the universe, it is more adequate
to use a fluid as the matter source. The Lagrangian of an ordinary perfect fluid \cite{Brown} can be written
in a Lorentz-invariant form:
\begin{equation}\label{LPF}
{\cal L}^{\rm PF}_{\mu\nu\rho\sigma}=-\epsilon_{\alpha\beta\gamma\delta}e^\alpha{}_\mu e^\beta{}_\nu
e^\gamma{}_\rho e^\delta{}_\sigma \rho+
\epsilon_{\alpha\beta\gamma\delta}J^\alpha e^\beta{}_\nu e^\gamma{}_\rho e^\delta{}_\sigma
\wedge\partial_\mu\phi,
\end{equation}
where $\phi$ is a scalar field, $J^\alpha$ is the particle number current which is Lorentz covariant and
satisfies $J^\alpha J_\alpha<0$, $\rho=\rho(n)$ is the energy density, and $n\equiv\sqrt{-J^\alpha
J_\alpha}$ is the particle number density. The Lagrangian (\ref{LPF}) is polynomial in the PGT variable
$e^\alpha{}_\mu$, but it is not polynomial in the DGT variables when it is reformulated into a dS-invariant
form, in which case the Lagrangian reads
\begin{eqnarray}\label{LPF2}
{\cal L}^{\rm PF}_{\mu\nu\rho\sigma}&=&-\epsilon_{ABCDE}(D_\mu\xi^A)(D_\nu\xi^B)(D_\rho\xi^C)(D_\sigma\xi^D)
(\xi^E/l)\,\rho\nonumber\\
&{}&+\epsilon_{ABCDE}J^A(D_\nu\xi^B)(D_\rho\xi^C)(D_\sigma\xi^D)\wedge(\xi^E/l)\,\partial_\mu\phi,
\end{eqnarray}
where $J^A$ is a dS-covariant particle number current, which satisfies $J^AJ_A<0$ and $J^A\xi_A=0$,
$\rho=\rho(n)$ and $n\equiv\sqrt{-J^AJ_A}$. Because $l^{-1}$ appears in Eq. (\ref{LPF2}), the Lagrangian
is not polynomial in $\xi^A$.

A straightforward way to modify Eq. (\ref{LPF2}) into a polynomial Lagrangian is to multiply it by $l$.
In the Lorentz gauges, $J^4=0$, and we may define the invariant $J^\mu\equiv J^\alpha e_\alpha{}^\mu$.
Then the modified Lagrangian ${\cal L}'^{PF}_{\mu\nu\rho\sigma}
=-e\epsilon_{\mu\nu\rho\sigma}\rho l+e\epsilon_{\mu'\nu\rho\sigma}J^{\mu'}\wedge l\cdot\partial_\mu\phi$.
It can be verified that this Lagrangian
violates the particle number conservation law $\nabla_\mu J^\mu=0$, where $\nabla_\mu$ is the linearly
covariant, metric-compatible and torsion-free derivative. To preserve the particle number conservation, we
may replace $l\cdot\partial_\mu\phi$ by $\partial_\mu(l\phi)$, and the corresponding dS-invariant Lagrangian is
\begin{eqnarray}\label{LDF}
{\cal L}^{\rm DF}_{\mu\nu\rho\sigma}&=&-\epsilon_{ABCDE}(D_\mu\xi^A)(D_\nu\xi^B)(D_\rho\xi^C)(D_\sigma\xi^D)
\,\xi^E\rho(n)\nonumber\\
&{}&+\epsilon_{ABCDE}J^A(D_\nu\xi^B)(D_\rho\xi^C)(D_\sigma\xi^D)\wedge\left(\frac14D_\mu\xi^E\cdot\phi
+\xi^E\partial_\mu\phi\right).
\end{eqnarray}
The perfect fluid depicted by the above Lagrangian is called the polynomial dS fluid, or dS fluid for
short. In the Lorentz gauges,
\begin{eqnarray}
{\cal L}^{\rm DF}_{\mu\nu\rho\sigma}&=&-e\epsilon_{\mu\nu\rho\sigma}\rho l+
\epsilon_{\alpha\beta\gamma\delta}J^\alpha e^\beta{}_\nu e^\gamma{}_\rho e^\delta{}_\sigma
\wedge(\partial_\mu l\cdot\phi+l\,\partial_\mu\phi)\nonumber\\
&=&-e\epsilon_{\mu\nu\rho\sigma}\rho l+e\epsilon_{\mu'\nu\rho\sigma}J^{\mu'}\wedge\partial_\mu(l\phi),
\end{eqnarray}
which is equivalent to Eq. (\ref{LPF}) when $l$ is a constant.

Define the Lagrangian function $\mathscr{L}_{\rm DF}$ by ${\cal L}^{\rm DF}_{\mu\nu\rho\sigma}
=\mathscr{L}_{\rm DF}\,e\epsilon_{\mu\nu\rho\sigma}$, then $\mathscr{L}_{\rm DF}=-\rho l
+J^\mu\partial_\mu(l\phi)$. To compare the polynomial dS fluid with the ordinary perfect fluid,
let us consider a general model with the Lagrangian function
\begin{equation}\label{Lm}
\mathscr{L}_{\rm m}=-\rho l^k+J^\mu\partial_\mu(l^k\phi),
\end{equation}
where $k\in\mathbb{R}$. When $k=0$, it describes the ordinary perfect fluid; when $k=1$, it describes
the polynomial dS fluid. The variation of $S_{\rm m}\equiv\int dx^4 e\mathscr{L}_{\rm m}$ with respect
to $\phi$ gives the particle number conservation law $\nabla_\mu J^\mu=0$. The variation with respect
to $J^\alpha$ yields $\partial_\mu(l^k\phi)=-\mu U_\mu l^k$, where $\mu\equiv d\rho/dn=(\rho+p)/n$ is
the chemical potential, $p=p(n)$ is the pressure, and $U^\mu\equiv J^\mu/n$ is the 4-velocity of the
fluid particle.
Making use of these results,
one may check that the on-shell Lagrangian function is equal to $pl^k$, and the variational derivatives
\begin{equation}
\delta S_{\rm m}/\delta l=-k\rho l^{k-1},
\end{equation}
\begin{equation}
\delta S_{\rm m}/\delta \Gamma^{\alpha\beta}{}_\mu=0,
\end{equation}
\begin{equation}
\delta S_{\rm m}/\delta e^\alpha{}_\mu=(\rho+p)l^kU^\mu U_\alpha+pl^ke_\alpha{}^\mu.
\end{equation}
It is seen that $\delta S_{\rm m}/\delta l=0$ for the ordinary perfect fluid, while $\delta S_{\rm m}/\delta l
=-\rho$ for the polynomial dS fluid.

Finally, it should be noted that the polynomial dS fluid does not support a signature change corresponding
to $\xi^A\xi_A$ varying from negative to positive. The reason is that when $\xi^A\xi_A<0$, there exists
no $J^A$ which satisfies $J^AJ_A<0$ and $J^A\xi_A=0$.

\section{Cosmological solutions}
\subsection{Field equations for the universe}\label{4.1}
In this section, the coupling system of the SW gravity and the fluid model (\ref{Lm}) will be analyzed in the
homogenous, isotropic, parity-invariant and spatially flat universe characterized by the following ansatz
\cite{Alexander}:
\begin{equation}\label{e}
e^0{}_\mu=d_\mu t,\quad e^i{}_\mu=a\,d_\mu x^i,
\end{equation}
\begin{equation}\label{S}
S^0{}_{\mu\nu}=0,\quad S^i{}_{\mu\nu}=b\,e^0{}_\mu\wedge e^i{}_\nu,
\end{equation}
where $a$ and $b$ are functions of the cosmic time $t$, and $i=1,2,3$. On account of Eqs.
(\ref{e})--(\ref{S}), the Lorentz connection $\Gamma^{\alpha\beta}{}_\mu$ and curvature
$R^{\alpha\beta}{}_{\mu\nu}$ can be calculated \cite{Alexander}. Further, assume that
$U^\mu=e_0{}^\mu$,
then $U_\mu=-e^0{}_\mu$, and so $U_\alpha=-\delta^0{}_\alpha$. Now the reduced form of each
term of Eqs. (\ref{eq-l})--(\ref{eq-e}) can be attained. In particular,
\begin{equation}
\epsilon_{\alpha\beta\gamma\delta}\,\epsilon^{\mu\nu\sigma\rho}e^{-1}R^{\alpha\beta}{}_{\mu\nu}
R^{\gamma\delta}{}_{\sigma\rho}=96(ha)^\cdot a^{-1}h^2,
\end{equation}
\begin{equation}
R=6[(ha)^\cdot a^{-1}+h^2],
\end{equation}
\begin{equation}
\epsilon_{0i\gamma\delta}\,\epsilon^{\mu\nu\sigma\rho}e^{-1}\partial_\nu l\cdot R^{\gamma\delta}
{}_{\sigma\rho}=-4h^2\dot{l}\,e_i{}^\mu,
\end{equation}
\begin{equation}
\epsilon_{ij\gamma\delta}\,\epsilon^{\mu\nu\sigma\rho}e^{-1}\partial_\nu l\cdot R^{\gamma\delta}
{}_{\sigma\rho}=0,
\end{equation}
\begin{equation}
T^\mu{}_{0i}=-2b\,e_i{}^\mu,\ \ T^\mu{}_{ij}=0,
\end{equation}
\begin{equation}
G^\mu{}_0=-3h^2e_0{}^\mu,
\end{equation}
\begin{equation}
G^\mu{}_i=-[2(ha)^\cdot a^{-1}+h^2]e_i{}^\mu,
\end{equation}
\begin{equation}
\delta S_{\rm m}/\delta e^0{}_\mu=-\rho l^ke_0{}^\mu,
\end{equation}
\begin{equation}
\delta S_{\rm m}/\delta e^i{}_\mu=pl^ke_i{}^\mu,
\end{equation}
where $\cdot$ on top of a quantity or being a superscript denotes the differentiation with respect to $t$,
and $h=\dot a/a-b$. Substitution
of the above equations into Eqs. (\ref{eq-l})--(\ref{eq-e}) leads to
\begin{equation}\label{cos-1}
(ha)^\cdot a^{-1}(h^2+l^{-2})+l^{-2}(h^2-\Lambda)=k\rho l^{k-1}/24\kappa,
\end{equation}
\begin{equation}\label{cos-2}
(h^2+l^{-2})\dot{l}-2b\,l^{-1}=0,
\end{equation}
\begin{equation}\label{cos-3}
8\kappa l^{-1}(-3h^2+\Lambda)=\rho l^k,
\end{equation}
\begin{equation}\label{cos-4}
8\kappa l^{-1}[-2(ha)^\cdot a^{-1}-h^2+\Lambda]=-pl^k,
\end{equation}
which constitute the field equations for the universe.

\subsection{Solutions for the field equations}\label{4.2}
Before solving the field equations (\ref{cos-1})--(\ref{cos-4}), let us first derive the continuity
equation from the field equations. Rewrite Eq. (\ref{cos-3}) as
\begin{equation}\label{h2}
h^2=l^{-2}-\rho l^{k+1}/24\kappa.
\end{equation}
Substituting Eq. (\ref{h2}) into Eq. (\ref{cos-4}) yields
\begin{equation}\label{ha}
(ha)^\cdot a^{-1}=l^{-2}+(\rho+3p)l^{k+1}/48\kappa.
\end{equation}
Multiply Eq. (\ref{ha}) by $2h$, making use of Eq. (\ref{h2}) and $h=\dot a/a-b$, one gets
\begin{equation}\label{hhd}
2h\dot{h}=(\rho+p)l^{k+1}\dot{a}a^{-1}/8\kappa-2b(ha)^\cdot a^{-1},
\end{equation}
in which, according to Eqs. (\ref{cos-1}), (\ref{cos-2}) and (\ref{h2}),
\begin{equation}\label{bha}
2b(ha)^\cdot a^{-1}=\dot{l}[(k+1)\rho l^k/24\kappa+2l^{-3}].
\end{equation}
Differentiate Eq. (\ref{h2}) with respect to $t$, and compare it with Eqs. (\ref{hhd})--(\ref{bha}),
one arrives at the continuity equation
\begin{equation}\label{ceq}
\dot{\rho}+3(\rho+p)\dot{a}a^{-1}=0,
\end{equation}
which is, unexpectedly, the same as the usual one. Suppose that $p=w\rho$, where $w$ is a constant.
Then Eq. (\ref{ceq}) has the solution
\begin{equation}\label{rho}
\rho=\rho_0(a/a_0)^{-3(1+w)},
\end{equation}
where $a_0$ and $\rho_0$ are the values of $a$ and $\rho$ at some moment $t_0$.

Now it is ready to solve Eqs. (\ref{cos-1})--(\ref{cos-3}), while Eq. (\ref{cos-4}) is replaced by
Eq. (\ref{ceq}) with the solution (\ref{rho}). Firstly, substitute Eqs. (\ref{h2})--(\ref{ha}) into Eq.
(\ref{cos-1}), one finds
\begin{equation}\label{rhol}
\rho l^{k+3}=48\kappa(3w-k-1)/(3w+1).
\end{equation}
Assume that $\kappa<0$, then according to the above relation, $\rho l^{k+3}>0$ implies
$(3w-k-1)/(3w+1)<0$. We only concern the cases with $k=0,1$, and so assume that $k+1>-1$,
then $\rho l^{k+3}>0$ constrains $w$ by
\begin{equation}\label{w}
-\frac1 3<w<\frac{k+1} 3.
\end{equation}
For the ordinary fluid ($k=0$), the pure radiation ($w=1/3$) cannot exist. In fact, on account
of Eq. (\ref{rhol}), $\rho l^3=0$ in this case, which is unreasonable. This problem is similar
to that appeared in Ref. \cite{Alexander}. On the other hand, for the dS fluid ($k=1$),
Eq. (\ref{w})
becomes $-1/3<w<2/3$, which contains both the cases with pure matter ($w=0$) and pure radiation
($w=1/3$). Generally, the combination of Eqs. (\ref{rho}) and (\ref{rhol}) yields
\begin{equation}\label{l}
l=l_0(a/a_0)^{\frac{3(w+1)}{k+3}},
\end{equation}
where $l_0$ is the value of $l$ when $t=t_0$, and is related to $\rho_0$ by Eq. (\ref{rhol}).

Secondly, substitute Eq. (\ref{h2}) into Eq. (\ref{cos-2}), and utilize Eqs. (\ref{rhol})
and (\ref{l}), one gets
\begin{equation}\label{b-H}
b=\frac{3(w+1)(k+2)}{(3w+1)(k+3)}\dot a a^{-1},
\end{equation}
and hence
\begin{equation}\label{h}
h=\frac{3w-2k-3}{(3w+1)(k+3)}\dot a a^{-1}.
\end{equation}

Thirdly, substitution of Eqs. (\ref{rhol}) and (\ref{h}) into Eq. (\ref{cos-3}) leads to
\begin{equation}\label{H}
\dot{a}a^{-1}=H_0(l_0/l),
\end{equation}
where $H_0\equiv(\dot a a^{-1})_{t_0}$ is the Hubble constant, being related to $l_0$ by
\begin{equation}\label{H0}
H_0=\sqrt{\frac{3w+1}{-3w+2k+3}}\cdot(k+3)l_0^{-1}.
\end{equation}
Here note that Eq. (\ref{w}) implies that $3w+1>0$, $-3w+k+1>0$, $k+1>-1$, and so $-3w+2k+3>0$.
In virtue of Eqs. (\ref{b-H}), (\ref{H}) and (\ref{l}), one has
\begin{equation}\label{b}
b=b_0(a_0/a)^{\frac{3(w+1)}{k+3}},
\end{equation}
where $b_0$ is related to $H_0$ by Eq. (\ref{b-H}). Moreover, substitute Eq. (\ref{l}) into
Eq. (\ref{H}) and solve the resulting equation, one attains
\begin{equation}\label{a}
(a/a_0)^{\frac{3(w+1)}{k+3}}-1=\frac{3(w+1)}{k+3}\cdot H_0(t-t_0).
\end{equation}

In conclusion, the solutions for the field equations (\ref{cos-1})--(\ref{cos-4}) are given by
Eqs. (\ref{rho}), (\ref{l}), (\ref{b}) and (\ref{a}), with the independent constants $a_0$, $H_0$
and $t_0$.

\subsection{Comparison with observations}\label{4.3}
If $k$ is specified, we can determine the value of the coupling constant $\kappa$ from the observed
values of $H_0=67.4\,{\rm km}\cdot{\rm s}^{-1}\cdot{\rm Mpc}^{-1}$ and $\Omega_0\equiv8\pi\rho_0/3H_0^2
=0.315$ \cite{Planck}. For example, put $k=1$, then according to Eq. (\ref{H0}) (with $w=0$), one has
\begin{equation}\label{l0}
l_0=4/\sqrt{5}H_0=8.19\times10^{17}\,{\rm s}.
\end{equation}
Substitution of Eq. (\ref{l0}) and $\rho_0=3H_0^2\Omega_0/8\pi=1.79\times10^{-37}\,{\rm s}^{-2}$ into
Eq. (\ref{rhol}) yields
\begin{equation}
\kappa=-\rho_0l_0^4/96=-8.41\times10^{32}\,{\rm s}^2.
\end{equation}
This value is an important reference for the future work which will explore the viability of the model
in the solar system scale.

Also, we can compare the deceleration parameter $q\equiv-a\ddot{a}/\dot{a}^2$ derived from the above
models with the observed one. With the help of Eqs. (\ref{H}) and (\ref{l}), one finds $\dot{a}\sim
a^{(k-3w)/(k+3)}$, then $\ddot{a}=\frac{k-3w}{k+3}\cdot\dot{a}^2a^{-1}$, and so
\begin{equation}\label{q}
q=\frac{3w-k}{k+3}.
\end{equation}
Put $w=0$, it is seen that the universe accelerates ($q<0$) if $k>0$, linearly expands ($q=0$) if
$k=0$, and decelerates ($q>0$) if $k<0$. In particular, for the model with an ordinary fluid
($k=0$), the universe expands linearly\,\footnote{This result is different from that in Ref.
\cite{Alexander}, where the cosmological solution describes a decelerating universe. It shows that the
SW model is not equivalent to the model considered in Ref. \cite{Alexander}.}; while for the model with
a dS fluid ($k=1$), the universe accelerates with $q=-1/4$, which is consistent with the
observational result $-1\leq q_0<0$ \cite{Riess,Schmidt,Perlmutter}, where $q_0$ is the present-day value
of $q$. It should be noted that Eq. (\ref{q}) implies that $q$ is a constant when $w$ is a constant, and so
the models cannot describe the transition from deceleration to acceleration when $w$ is a constant.

\section{Remarks}\label{5}
It is shown that the requirement of regular Lagrangian may be crucial for DGT, as it is shown that the
SW gravity coupled with an ordinary perfect fluid (whose Lagrangian is not regular with respect to $\xi^A$
when $\xi^A\xi_A=0$) does not permit a radiation epoch and the acceleration of the universe, while the SW
gravity coupled with a polynomial dS fluid (whose Lagrangian is regular with respect to $\xi^A$) is out of
these problems. Yet, the latter model is still not a realistic model, because it cannot
describe the transition from deceleration to acceleration in the matter epoch.

There are two possible ways to find a more reasonable model. The first is to modify the gravitational
part to be the general quadratic model (\ref{LG}), which is a special case of the at most quadratic
model proposed in Refs. \cite{Westman12,Westman14}, but the coupling of which with the polynomial dS fluid is
unexplored. It is
unknown whether the effect of the $\kappa_2$ term could solve the problem encountered in the SW gravity.

The second way is to modify the matter part. Although the Lagrangian of the polynomial dS fluid is regular with
respect to
$\xi^A$, it is not regular with respect to $J^A$ when $\xi^A\xi_A=0$, in which case there should be $J^A
J_A\geq0$, and so the number density $n\equiv\sqrt{-J^A J_A}$ is not regular.
Maybe one could find a new fluid model whose Lagrangian is regular with respect to all the
variables, based on the polynomial models for fundamental fields proposed in Refs. \cite{Pagels,Westman13}.

\section*{Acknowledgments}
I thank Profs. S.-D. Liang and Z.-B. Li for their abiding help. Also, I would to thank my parents
and my wife. This research is supported by the National Natural Science Foundation for Young Scientists
of China under Grant No. 12005307.

\end{document}